\documentclass[11pt,a4paper]{article}
\usepackage{flafter}

\usepackage{latexsym}
\usepackage[fleqn]{amsmath}
\usepackage{amsfonts,amssymb}

\newtheorem{theorem}{Theorem}  
\newtheorem{lemma}{Lemma}
\newtheorem{definition}{Definition}   
\newenvironment{proof}{\begin{trivlist}
 \item[\hspace{\labelsep}{\em\noindent Proof.~}]
 }{\qed\end{trivlist}} 

\newcommand{\eps}{\varepsilon}
\newcommand{\ket}[1]{\mbox{$| #1 \rangle$}}

\newcommand{\braket}[1]{\mbox{$\langle #1 | #1 \rangle$}}

\newcommand{\integer}{{\mathbb Z}}
\newcommand{\reals}{{\mathbb R}}

\def\squareforqed{\hbox{\rlap{$\sqcap$}$\sqcup$}}
\def\qed{\ifmmode\squareforqed\else{\unskip\nobreak\hfil
\penalty50\hskip1em\null\nobreak\hfil\squareforqed
\parfillskip=0pt\finalhyphendemerits=0\endgraf}\fi}

\newenvironment{algorithm}[1]{\medskip\noindent%
\itemsep0pt\begin{trivlist}\item[]%
{\flushleft\textbf{Algorithm( #1 )}}}%
{\end{trivlist}\medskip}

\begin{document}

\title{Robust Quantum Algorithms with $\eps$-Biased Oracles}

\author{
Tomoya Suzuki \and
Shigeru Yamashita \and
Masaki Nakanishi \and
Katsumasa Watanabe
\\
Graduate School of Information Science, Nara Institute of Science and Technology\\
{\tt \{tomoya-s, ger, m-naka, watanabe\}@is.naist.jp}
}

\date{}

\maketitle

\begin{abstract}
This paper considers the quantum query complexity of {\it $\eps$-biased
 oracles} that return the correct value with probability only $1/2 + \eps$.
In particular, we show a quantum algorithm to compute
$N$-bit OR functions with $O(\sqrt{N}/{\eps})$ queries to $\eps$-biased
oracles. This improves the known upper bound of $O(\sqrt{N}/{\eps}^2)$
and matches the known lower bound; we answer the conjecture raised by
the paper~\cite{yamashita} affirmatively.  
We also show a quantum algorithm to cope with the situation in which we
have no knowledge about the value of $\eps$. This contrasts with the
corresponding classical situation, where it is almost hopeless to 
achieve more than a constant success probability without knowing the value of
$\eps$.  
\end{abstract}

\section{Introduction} 

Quantum computation has attracted much attention  
since Shor's celebrated quantum algorithm for factoring large
integers~\cite{Shor94} and Grover's quantum search algorithm~\cite{grover}. 
One of the central issues in this research field has been the {\it
  quantum query complexity}, where we are interested in both upper and
lower bounds of a necessary number of oracle calls to solve certain 
problems~\cite{AA03,ambainis,Amb03b,Beals98,BBHT96,BDHHMSW00,Shi02}.
In these studies, oracles are assumed to 
be {\it perfect}, i.e., they return the correct value with certainty. 

In the classical case, there have been many studies
(e.g.,~\cite{noisy}) that discuss the case of when oracles are {\it
  imperfect} (or often called {\it noisy}), i.e., they may return
incorrect answers.    
In the quantum setting,  H{\o}yer et al.~\cite{mosca} proposed an 
excellent quantum algorithm, which we call the {\it robust quantum search
  algorithm} hereafter, to compute the OR function of $N$ values,
each of which can be accessed through
a quantum ``imperfect'' oracle. Their quantum ``imperfect'' 
 oracle can be described as follows: When the
 content of the query register is  $x$ ($1 \leq x \leq N$), the
 oracle returns a  quantum pure state  from which we can measure the
 correct value of  $f(x)$ with a constant probability. This noise
 model naturally fits into  quantum subroutines with errors. (Note
 that most existing quantum algorithms have some errors.)  
More precisely,  their algorithm robustly computes $N$-bit OR
functions with $O(\sqrt{N})$ queries to an imperfect oracle, which is
only a constant factor worse than the perfect oracle case. Thus, they
claim that their 
algorithm does not need a serious overhead to cope with the
imperfectness of the oracles. 
Their method has been extended to a robust quantum algorithm to output 
all the $N$ bits by using $O(N)$  queries~\cite{rparity} by Buhrman
et~al. This obviously implies that $O(N)$ queries are  
enough to compute the parity of the $N$ bits, which contrasts with the
classical $\Omega(N\log N)$ lower bound given in~\cite{noisy}.  

It should be noted that, in the classical setting, we do not need an
overhead to compute OR functions with imperfect oracles either, i.e.,
$O(N)$ queries are enough to compute $N$-bit OR functions  even if
an oracle is imperfect \cite{noisy}.  Nevertheless, 
the robust quantum search algorithm by H{\o}yer et al.~\cite{mosca} implies
that we can still enjoy the quadratic speed-up of the quantum search
when computing OR functions, even in the imperfect oracle case, i.e.,  
$O(\sqrt{N})$ vs. $O(N)$.  
However, this is not true when we consider the probability of getting  
the correct value from the imperfect oracles {\it explicitly}  by 
using the following model: When the query register
is $x$,  the oracle returns a quantum pure state from which we can
measure the correct value of $f(x)$ with  probability $1/2 + {\eps}_x$,
where we assume
$\eps \leq \eps_x$
for any $x$ and we know the value of $\eps$. In this paper, we call this
imperfect quantum oracle {\it an ${\eps}$-biased oracle} (or a biased
oracle for short) by following the paper~\cite{yamashita}.   
Then, the precise query complexity of the above robust quantum search 
algorithm to compute OR functions with an ${\eps}$-biased oracle can
be rewritten as $O(\sqrt{N}/{\eps}^2)$, which can also be found in~\cite{rparity}.   
For the same problem, we need $O(N/{\eps}^2)$ queries in the classical
setting since $O(1/{\eps}^2)$ instances of majority voting of the output   
of an ${\eps}$-biased oracle is enough to boost the success
probability to some constant value.
This means that the above robust quantum search algorithm does not achieve
the quadratic speed-up anymore if we consider the error probability
explicitly.   

Adcock et al.~\cite{gl} first considered the error probability explicitly 
in the quantum oracles, then Iwama et al.~\cite{yamashita} continued
to study ${\eps}$-biased oracles: they show the lower bound of
computing OR is $\Omega(\sqrt{N}/{\eps})$ and the matching upper bound
when ${\eps}_x$ are the same for all $x$. Unfortunately, this restriction
to oracles obviously cannot be applied in general. Therefore, for the 
general biased oracles, there have been a gap between the lower and
upper bounds although the paper~\cite{yamashita} conjectures that they
should match at $\Theta(\sqrt{N}/{\eps})$.   

{\bf Our Contribution.~} 
In this paper, we show that the robust quantum search can be done with
$O(\sqrt{N}/{\eps})$ queries. Thus, we answer the conjecture raised by
the paper~\cite{yamashita} affirmatively, meaning that we can
still  enjoy the quantum quadratic speed-up to compute OR functions
even when we consider the error probability explicitly.   
The overhead factor of $1/{\eps}^2$ in the complexity of the original
robust quantum search (i.e., $O(\sqrt{N}/{\eps}^2)$) essentially 
comes from the classical majority voting in their recursive algorithm.  
Thus, our basic strategy is to utilize {\it quantum amplitude 
  amplification and estimation}~\cite{qaa} instead of majority
voting to boost the success probability to some constant value. 
This overall strategy is an extension of the idea in the 
paper~\cite{yamashita}, but we carefully perform the quantum amplitude 
amplification and estimation in quantum parallelism with appropriate
accuracy to avoid the above-mentioned restriction to oracles assumed
in~\cite{yamashita}.  

In most existing (classical and quantum) algorithms with imperfect 
oracles, it is implicitly assumed that we know the value of 
$\eps$. Otherwise, it seems impossible to know when we can
stop the trial of majority voting with a guarantee of a more than
constant success probability of the whole algorithm. 
However, we show that, in the quantum setting, we can construct a robust
algorithm even when $\eps$ is unknown. More precisely, we can estimate 
unknown $\eps$ with appropriate accuracy, which then can be used to
construct robust quantum algorithms. Our estimation algorithm also
utilizes quantum amplitude estimation, thus it can be considered as an
interesting application of quantum amplitude amplification, which
seems to be impossible in the classical setting.    

\section{Preliminaries}
In this section, we introduce the quantum computing and the query complexity.
We also define quantum biased oracles.

\subsection{Quantum State and Evolution}
A state of $n$-\textit{qubit} quantum register $\ket{\psi}$
is a superposition of $2^n$ classical strings with length $n$, i.e.,
$\ket{\psi} = \sum_{x}\alpha_x\ket{x}$
where $x\in\{0,1\}^n$ and
the \textit{amplitude}s $\alpha_x$ are complex numbers consistent with the normalization condition: 
$\sum_x|\alpha_x|^2 = 1$. If we \textit{measure} the state $\ket{\psi}$ with respect to the standard basis,
we observe $\ket{x}$ with probability $|\alpha_x|^2$ and after the measurement
the state $\ket{\psi}$ collapses into $\ket{x}$.

Without measurements, a quantum system can evolve satisfying
the normalization condition. These evolutions are represented by unitary transformations.
The following \textit{Fourier transform} is a famous example that acts on several qubits.
\begin{definition}
  For any integer $M \geq 1$, a quantum Fourier transform ${\mathbf F}_M$ is defined by
  \[
    {\mathbf F}_M \;:\; \ket{x} \;\longmapsto\;
    \frac{1}{\sqrt M} \sum_{y=0}^{M-1}
    e^{2 \pi \imath x y/M} \ket{y} ~~ (0\le x <M).
    \]
\end{definition}
In this paper, unitary transformations controlled by
other registers are often used.
For example, one of them acts as some unitary transformation if the control qubit is $\ket{1}$,
otherwise it acts as identity. 
The following operator $\Lambda_M$ is also one of their applications.
\begin{definition}
  For any integer $M \geq 1$ and any unitary operator $\mathbf U$, the operator $\Lambda_M(\mathbf U)$ is defined by
  \begin{equation*}
    \ket{j}\ket{y} \;\longmapsto\;\begin{cases}
    \ket{j}\mathbf{U}^j\ket{y}  & (0 \leq j < M)\\
    \ket{j}\mathbf{U}^M\ket{y}  & (j \geq M).
    \end{cases}
  \end{equation*}
  $\Lambda_M$ is controlled by the first register $\ket{j}$ in this case.
  $\Lambda_M(\mathbf U)$ uses $\mathbf U$ for $M$ times.
\end{definition}
It is also known that quantum transformations can compute all classical functions.
Let $g$ be any classically computable function with $m$ input and $k$ output bits.
Then, there exists a unitary transformation $\mathbf U_g$ corresponding to the computation of $g$:
for any $x \in \{0,1\}^m$ and $y \in \{0,1\}^k$,
$\mathbf U_g$ maps $\ket{x}\ket{y}$ to $\ket{x}\ket{y\oplus g(x)}$, where
$\oplus$ denotes the bitwise exclusive-OR.

\subsection{Query Complexity}
In this paper, we are interested in the query complexity,
which is discussed in the following model.
Suppose we want to compute some function $\mathcal F$ with an $N$-bit input
and
we can access each bit only through a given oracle $O$.
The query complexity is the number of queries to the oracle.
A quantum algorithm with $T$ queries is a sequence of unitary transformations:
$
U_0 \to O_1 \to U_1 \to \ldots \to O_T \to U_T,
$
where $O_i$ denotes the unitary transformation corresponding to the $i$-th query to the oracle $O$,
and $U_i$ denotes an arbitrary unitary transformation independent of the oracle.
Our natural goal is to find an algorithm to compute $\mathcal F$ with sufficiently large probability and with
the smallest number of oracle calls.

The most natural quantum oracles are quantum perfect oracles $O_f$ that
map $\ket{x}\ket{0^{m-1}}\ket{0}$ to $\ket{x}\ket{0^{m-1}}\ket{f(x)}$ for any $x\in[N]$.
Here, $\ket{0^{m-1}}$ is a work register that is always cleared before and after querying oracles.
On the other hand, quantum biased oracles, which we deal with in this paper, are defined as follows.

\begin{definition}\label{def:biased_oracle}
  A quantum oracle of a Boolean function $f$ with bias $\eps$ is a unitary
  transformation $O^\eps_f$ or its inverse ${O^\eps_f}^\dagger$ such that
  \begin{equation*}\label{definition}
    O^{\eps}_{f}\ket{x}\ket{0^{m-1}}\ket{0} =
    \ket{x} (\alpha_x\ket{w_x}\ket{f(x)}+\beta_x\ket{w'_x}\ket{\overline{f(x)}}),
  \end{equation*}
  where $|\alpha_x|^2 = 1/2 + \eps_x \geq 1/2 + \eps$ for any $x \in [N]$.
  Let also $\displaystyle \eps_{\mathrm{min}} = \min_x \eps_x$.
\end{definition}
Note that $0 < \eps \leq \eps_{\mathrm{min}} \leq \eps_x \leq 1/2$ for any $x$.
In practice, $\eps$ is usually given in some way and $\eps_{\mathrm{min}}$ or $\eps_{x}$ may be unknown.
Unless otherwise stated, we discuss the query complexity 
with a given biased oracle $O_f^\eps$ in the rest of the paper.

We can also consider phase flip oracles instead of the above-defined bit flip oracles.
A (perfect) phase flip oracle is defined as a map: $\ket{x}\ket{0^{m-1}}
\longmapsto (-1)^{f(x)}\ket{x}\ket{0^{m-1}}$, which is equivalent to the corresponding bit flip oracle $O_f$
in the perfect case,
since either oracle can be easily simulated by the other oracle with a pair of Hadmard gates.
In a biased case, however, the two oracles cannot always be converted to each other.
We need to take care of interference of the work registers, i.e., $ \ket{w_x}$ and $\ket{w'_x}$,
which are dealt with carefully in our algorithm.

\subsection{Amplitude Amplification and Estimation}
We briefly introduce a few known quantum algorithms often used in the following sections.
In \cite{qaa}, Brassard et al. presented amplitude amplification as follows.

\begin{theorem}\label{th:amplification}
  Let $\mathcal A$ be any quantum algorithm that uses no measurements and $\chi : \integer \to \{0,1\}$
  be any Boolean function that distinguishes between success or fail (good or bad).
  There exists a quantum algorithm that
  given the initial success probability $p>0$ of $\mathcal A$,
  finds a good solution with certainty
  using a number of applications of $\mathcal A$ and $\mathcal A^{-1}$, which
  is in $O(\frac{1}{\sqrt{p}})$ in the worst case.
\end{theorem}
In the amplitude amplification, a unitary operator
$\mathbf Q = -{\mathcal A}{\mathbf S_0}{\mathcal A^{-1}}{\mathbf S_\chi}$
is used. Here, $\mathbf S_0$ denotes an operator to flip the sign of amplitude
of the state $\ket{\mathbf 0}$,
and $\mathbf S_\chi$ denotes an operator to flip the signs of amplitudes of all
the good states. Applying $\mathbf Q$ to the state $\mathcal A\ket{\mathbf 0}$ for $j$ times, we have
\[
{\mathbf Q}^j {\mathcal A} \ket{\mathbf 0}
= \frac{1}{\sqrt p} \sin((2j+1)\theta_p) \,\ket{\Psi_1} \nonumber
+ \frac{1}{\sqrt{1-p}} \cos((2j+1)\theta_p) \,\ket{\Psi_0},
\]
where $\ket{\Psi_1}$ has all the good states, and $\braket{\Psi_1} = p = \sin^2(\theta_p)$ and $\ket{\Psi_1}$
is orthogonal to $\ket{\Psi_0}$.
After applying $\mathbf Q$ for about $\pi/{4\theta_p} \in O(1/\sqrt{p})$ times,
we can measure a good solution with probability close to $1$.
Note that we need to know information about the value of $p$ in some way to do so.
See \cite{qaa} for more details.

Brassard et al. also presented amplitude estimation in~\cite{qaa}.
We rewrite it in terms of phase estimation as follows.

\begin{theorem}\label{th:estimation}
  Let $\mathcal A,\chi$ and $p$ be as in Theorem~\ref{th:amplification} and $\theta_p = \sin^{-1}(\sqrt{p})$ such that
  $0 \le \theta_p \le \pi/2$.
  There exists a quantum algorithm $Est\_Phase(\mathcal A,\chi,M)$ that outputs
  $\tilde\theta_p$
  such that
  $|\theta_p - \tilde\theta_p| \leq \frac{\pi}{M}$,
  with probability at least ${8}/{\pi^2}$.
  It uses exactly $M$ invocations of $\mathcal A$ and $\chi$, respectively.
  If $\theta_p = 0$ then $\tilde\theta_p = 0$ with certainty,
  and if $\theta_p = {\pi}/{2}$ and $M$ is even, then $\tilde\theta_p = {\pi}/{2}$ with certainty.
\end{theorem}

\section{Computing OR with $\eps$-Biased Oracles}\label{se:known}
In this section, we assume that we have information about bias rate of the given biased oracle:
a value of $\eps$ such that $0 < \eps \le \eps_{\mathrm{min}}$.
Under this assumption, in Theorem \ref{th:or}
we show that $N$-bit OR functions can be computed by using $O(\sqrt{N}/\eps)$ queries
to the given oracle $O_f^\eps$. Moreover, when we know $\eps_{\mathrm{min}}$, we can
present an optimal algorithm to compute OR with $O_f^\eps$.
Before describing the main theorem, we present the following key lemma.

\begin{lemma}\label{le:main}
  There exists a quantum algorithm that simulates a single query to an oracle $O^{1/6}_{f}$
  by using $O(1/\eps)$ queries to $O^{\eps}_{f}$ if we know $\eps$.
\end{lemma}
To prove the lemma, we replace
the given oracle~$O^\eps_f$ with a new oracle~$\tilde{O}^{\eps}_{f}$ for our convenience.
The next lemma describes the oracle~$\tilde{O}^{\eps}_{f}$ and how to construct it from $O^{\eps}_{f}$.

\begin{lemma}
  There exists a quantum oracle $\tilde{O}^{\eps}_{f}$ that consists of one
  $O^{\eps}_{f}$ and one ${O^{\eps}_{f}}^{\dagger}$ such that for any $x \in [N]$
  \begin{equation}
  \tilde{O}^{\eps}_{f}\ket{x,0^m,0} = (-1)^{f(x)}2\eps_{x}\ket{x,0^m,0}+\ket{x,\psi_x}, \label{eq:oracle}
  \end{equation}
  where $\ket{x,\psi_x}$ is orthogonal to $\ket{x,0^m,0}$ and its norm is $\sqrt{1-4{\eps_x}^2}$.
\end{lemma}

\begin{proof}
  We can show the construction of $\tilde{O}^{\eps}_{f}$ in a similar way in Lemma~1 in~\cite{yamashita}.
\qed
\end{proof}

Now, we describe our approach to Lemma~\ref{le:main}. The oracle
$O^{1/6}_f$ is simulated by the given oracle $O^\eps_f$ based on the following idea.
According to \cite{yamashita},
if the query register $\ket{x}$ is not in a superposition,
phase flip oracles can be simulated with sufficiently large probability:
by using amplitude estimation through $\tilde O_f^\eps$, we can estimate the value of $\eps_x$,
then by using the estimated value and applying
amplitude amplification to the state in (\ref{eq:oracle}),
we can obtain the state $(-1)^{f(x)}\ket{x,0^m,0}$ with high probability.
In Lemma~\ref{le:main}, we essentially simulate the phase flip oracle by using
the above algorithm in a superposition of $\ket{x}$.
Note that we convert the phase flip oracle into the bit flip version in the lemma.

We will present the proof of Lemma~\ref{le:main} after the following lemma,
which shows that amplitude estimation can work in quantum parallelism.
$Est\_Phase$ in Theorem~\ref{th:estimation}
is straightforwardly extended to $Par\_Est\_Phase$ in Lemma~\ref{le:pestimation},
whose proof can be found in the Appendix.

\begin{lemma}\label{le:pestimation}
  Let $\chi:\integer \to \{0,1\}$ be any Boolean function,
  and let $\mathcal{O}$  be any quantum oracle that uses no measurements such that
  \[
  \mathcal{O}\ket{x}\ket{\mathbf{0}}
  = \ket{x}\mathcal{O}_x\ket{\mathbf{0}}
  = \ket{x}\ket{\Psi_x}
  = \ket{x}(\ket{\Psi_x^1} + \ket{\Psi_x^0}),
  \]
  where a state $\ket{\Psi_x}$ is divided into a good state $\ket{\Psi_x^1}$
  and a bad state $\ket{\Psi_x^0}$ by $\chi$.
  Let $\sin^2(\theta_x) = \braket{\Psi_x^1}$
  be the success probability of $\mathcal O_x\ket{\mathbf 0}$
  where $0 \leq \theta_x \leq {\pi}/{2}$.
  There exists a quantum algorithm $Par\_Est\_Phase(\mathcal{O},\chi,M)$
  that changes states as follows:
  \begin{equation*}
    \ket{x}\ket{\mathbf{0}}\ket{\mathbf{0}}
    \;\longmapsto\; \ket{x}\otimes
    \sum_{j=0}^{M-1}\delta_{x,j}\ket{v_{x,j}}\ket{\tilde{\theta}_{x,j}},
  \end{equation*}
  where $\displaystyle \sum_{j:|\theta_x-\tilde{\theta}_{x,j}|
    \leq \frac{\pi}{M}}|\delta_{x,j}|^2 \geq \frac{8}{\pi^2}$ \;
  for any $x$, and
  $\ket{v_{x,i}}$ and $\ket{v_{x,j}}$ are mutually orthonormal vectors for any $i, j$.
  It uses $\mathcal O$ and its inverse for $O(M)$ times.
\end{lemma}

\begin{proof} (of Lemma~\ref{le:main})
  
  We will show a quantum algorithm that changes states as follows:
  \[
  \ket{x}\ket{\mathbf 0}\ket{0}
  \; \longmapsto \;
  \ket{x} (\alpha_x\ket{w_x}\ket{f(x)}+\beta_x\ket{w'_x}\ket{\overline{f(x)}}),
  \]
  where $|\alpha_x|^2 \geq 2/3$ for any $x$, using $O(1/\eps)$ queries to $O^{\eps}_{f}$.
  The algorithm performs amplitude amplification following amplitude estimation
  in a superposition of $\ket{x}$.

  At first, we use amplitude estimation in parallel to estimate $\eps_x$ or to know how many times
  the following amplitude amplification procedures should be repeated.
  Let $\sin\theta = 2\eps$ and $\sin\theta_x = 2\eps_x$ such that $0<\theta,\theta_x\le\pi/2$.
  Note that $\Theta(\theta) = \Theta(\eps)$ since $\sin\theta\leq\theta\leq\frac{\pi}{2}\sin\theta$
  when $0\le\theta\le\pi/2$.
  Let also $M_1 = \left\lceil\frac{3\pi(\pi+1)}{\theta}\right\rceil$ and
  $\chi$ be a Boolean function that divides a state in (\ref{eq:oracle})
  into a good state $(-1)^{f(x)}2\eps_x\ket{0^{m+1}}$ and a bad state $\ket{\psi_x}$.
  The function $\chi$ checks only whether the state is $\ket{0^{m+1}}$ or not;
  therefore, it is implemented easily.
  By Lemma~\ref{le:pestimation},
  $Par\_Est\_Phase(\tilde{O}^{\eps}_{f}$,$\chi$,$M_1)$ maps
  \begin{equation*}
    \ket{x}\ket{\mathbf 0}\ket{\mathbf 0}\ket{\mathbf 0}
    \; \longmapsto \;
    \ket{x}\otimes\sum_{j=0}^{M-1}\delta_{x,j}\ket{v_{x,j}}\ket{\tilde{\theta}_{x,j}}\ket{\mathbf 0}, \label{pestimated}
  \end{equation*}
  where $\displaystyle \sum_{j:|\theta_x-\tilde{\theta}_{x,j}|\leq
    \frac{\theta}{3(\pi+1)}}|\delta_{x,j}|^2 \geq \frac{8}{\pi^2}$ \; for any $x$,
  and
  $\ket{v_{x,i}}$ and $\ket{v_{x,j}}$ are mutually orthonormal vectors for any $i, j$.
  This state has the good estimations of $\theta_x$ in the third register with high probability.
  The fourth register $\ket{\mathbf 0}$ remains large enough to perform the following steps.

  The remaining steps basically perform amplitude amplification
  by using the estimated values $\tilde\theta_{x,j}$,
  which can realize a phase flip oracle.
  Note that in the following steps a pair of Hadmard transformations
  are used to convert the phase flip oracle into our targeted oracle.
  
  Based on the de-randomization idea as in~\cite{yamashita}, we calculate $m^*_{x,j} =
  \left\lceil \frac{1}{2}\left(\frac{\pi}{2\tilde\theta_{x,j}} - 1\right) \right\rceil$,
  $\theta_{x,j}^* = \frac{\pi}{4m^*_{x,j}+2}$, $p_{x,j}^* = \sin^2(\theta_{x,j}^*)$ and
  $\tilde{p}_{x,j} = \sin^2(\tilde\theta_{x,j})$ in the superposition, and apply an Hadmard transformation
  to the last qubit.
  Thus we have
  \begin{eqnarray*}
  \ket{x}
  \biggl(
  \sum_{j=0}^{M-1}\delta_{x,j}\ket{v_{x,j}}\ket{\tilde{\theta}_{x,j}}
  \ket{m^*_{x,j}}\ket{\theta^*_{x,j}}\ket{p^*_{x,j}}\ket{\tilde{p}_{x,j}}
  \otimes
  \ket{0^{m+1}}
  \ket{0}
  \otimes{\frac{1}{\sqrt{2}}}\left(\ket{0}+\ket{1}\right)
  \biggr).
  \end{eqnarray*}
  Next, let
  $\mathbf R : \ket{p^*_{x,j}}\ket{\tilde{p}_{x,j}}\ket{0} \to
  \ket{p^*_{x,j}}\ket{\tilde{p}_{x,j}}
  \left(\sqrt{\frac{p^*_{x,j}}{\tilde{p}_{x,j}}}\ket{0}+\sqrt{1-\frac{p^*_{x,j}}{\tilde{p}_{x,j}}}\ket{1}\right)$
  be a rotation
  and let $\mathbf O = \tilde O_f^\eps \otimes \mathbf R$ be a new oracle.
  We apply $\mathbf O$ followed by $\Lambda_{M_2}(\mathbf Q)$, where
  $M_2 = \left\lceil \frac{1}{2}\left(\frac{3\pi(\pi+1)}{2(3\pi+2)\theta} + 1\right) \right\rceil$
  and $\mathbf Q = -\mathbf O (\mathbf I \otimes \mathbf{S}_0) \mathbf O^{-1} (\mathbf I \otimes \mathbf{S}_\chi) $;
  $\mathbf S_0$ and $\mathbf S_\chi$ are defined appropriately.
  $\Lambda_{M_2}$ is controlled by the register $\ket{m_{x,j}^*}$, and 
  $\mathbf Q$ is applied to the registers $\ket{x}$ and $\ket{0^{m+1}}\ket{0}$ if the last qubit is $\ket{1}$.
  Let $\mathbf O_x$ denote the unitary operator
  such that $\mathbf O\ket{x}\ket{0^{m+1}}\ket{0} = \ket{x}\mathbf O_x\ket{0^{m+1}}\ket{0}$.
  Then we have the state (From here, we write only the last three registers.)
  \begin{equation}
    \sum_{j=0}^{M-1}
    \frac{\delta_{x,j}}{\sqrt{2}}
    \left(
    \ket{0^{m+1}}\ket{0}\ket{0}
    +\mathbf Q_x^{{m}_{x,j}}\mathbf O_x\left(\ket{0^{m+1}}\ket{0}\right)\ket{1}
    \right)\label{eq:flip},
  \end{equation}
  where
  $\mathbf Q_x = - \mathbf O_x \mathbf S_0 \mathbf O_x^{-1} \mathbf S_\chi$
  and
  $m_{x,j} = \min(m_{x,j}^*,M_2)$
  for any $x,j$.
  We will show that the phase flip oracle is simulated if
  the third register $\ket{\tilde\theta_{x,j}}$ has the good estimation of $\theta_x$ and
  the last register has $\ket{1}$.
  Equation~(\ref{eq:flip}) can be rewritten as
  \begin{eqnarray*}
  \sum_{j=0}^{M-1}
  \frac{\delta_{x,j}}{\sqrt{2}}
  \biggl(
  \ket{0^{m+1},0}\ket{0}
  + \left((-1)^{f(x)}\gamma_{x,j}\ket{0^{m+1},0} + \ket{\varphi_{x,j}}\right) \ket{1}
  \biggr),
  \end{eqnarray*}
  where $\ket{\varphi_{x,j}}$ is orthogonal to $\ket{0^{m+1},0}$ and its norm is $\sqrt{1-\gamma_{x,j}^2}$.
  Suppose that the third register has $\ket{\tilde\theta_{x,j}}$
  such that $|\theta_x - \tilde\theta_{x,j}| \leq \frac{\theta_x}{3(\pi+1)}$.
  It can be seen that $m_{x,j} \le M_2$ if $|\theta_x - \tilde\theta_{x,j}| \leq \frac{\theta_x}{3(\pi+1)}$.
  Therefore, $\mathbf Q_x$ is applied for $m_{x,j}^*$ times, i.e., the number specified by the fourth register.
  Like the analysis of Lemma~2 in \cite{yamashita},
  it is shown that $\gamma_{x,j} \geq \sqrt{1-\frac{1}{9}}$.

  Finally, applying an Hadmard transformation to the last qubit again, we have the state
  \begin{eqnarray*}
    \sum_{j=0}^{M-1}
    {\frac{\delta_{x,j}}{2}}\Biggl(
    &&(1+(-1)^{f(x)}\gamma_{x,j})\ket{0^{m+2}}\ket{0}\\
    && \mbox{} +
    (1-(-1)^{f(x)}\gamma_{x,j})\ket{0^{m+2}}\ket{1}
    +
    \ket{\varphi_{x,j}}(\ket{0}-\ket{1})
    \Biggr).
  \end{eqnarray*}
  If we measure the last qubit, we have \ket{f(x)} with probability
  \begin{eqnarray*}
    &&\sum_{j=0}^{M-1} \left(\left|\frac{\delta_{x,j}(1+\gamma_{x,j})}{2}\right|^2 +
    \left|\frac{\delta_{x,j}\sqrt{1-\gamma_{x,j}^2}}{2}\right|^2\right)\\
    & \geq & \frac{1}{2}\sum_{j:|\theta_x - \tilde\theta_{x,j}|\leq\frac{\theta}{3(\pi+1)}}
    |\delta_{x,j}|^2\left(1+\gamma_{x,j}
    \right)  ~ \geq ~ \frac{2}{3}.
  \end{eqnarray*}
  Thus, the final quantum state can be rewritten as
  $\ket{x} (\alpha_x\ket{w_x}\ket{f(x)}+\beta_x\ket{w'_x}\ket{\overline{f(x)}}$,
  where $|\alpha_x|^2 \geq 2/3$ for any $x$.

  The query complexity of this algorithm is the cost of amplitude estimation $M_1$
  and amplitude amplification $M_2$, thus a total number of queries is $O(\frac{1}{\theta}) = O(\frac{1}{\eps})$.
  Therefore, we can simulate a single query to $O_f^{1/6}$
  using $O(\frac{1}{\eps})$ queries to $O_f^{\eps}$.
\qed
\end{proof}

Now, we describe the main theorem to compute OR functions with quantum biased oracles.

\begin{theorem}\label{th:or}
  There exists a quantum algorithm to compute $N$-bit OR
  with probability at least $2/3$
  using $O({\sqrt{N}}/{\eps})$ queries to a given oracle $O^{\eps}_{f}$ if we know $\eps$.
  Moreover, if we know $\eps_{\mathrm{min}}$, the algorithm uses $\Theta({\sqrt{N}}/{\eps_{\mathrm{min}}})$ queries.
\end{theorem}

\begin{proof}
  The upper bound $O({\sqrt{N}}/{\eps})$ is obtained by Lemma~\ref{le:main} and \cite{mosca}.
  In~\cite{mosca}, we can see an algorithm to compute OR with probability at least $2/3$
  using $O(\sqrt{N})$ queries
  to an oracle $O^{1/6}_f$. When an oracle $O^\eps_f$ and a value of $\eps$ are given, we can simulate one query
  to an oracle $O^{1/6}_f$ using $O(1/\eps)$ queries to $O^\eps_f$ by Lemma~\ref{le:main}.
  Therefore, we can compute OR using $O({\sqrt{N}}/{\eps})$ queries to an oracle $O^\eps_f$.

  The lower bound $\Omega({\sqrt{N}}/{\eps_{\mathrm{min}}})$ is also obtained by Theorem~6 in \cite{yamashita}.
  The theorem states that for any problem, if the lower bound $\Omega(T)$ can be shown by Ambainis' method in
  the noiseless case, then the lower bound $\Omega(T/\eps_{\mathrm{min}})$ can also be shown in the noisy case.
  For computing $N$-bit OR functions, $\Omega(\sqrt{N})$ can be shown by Ambainis' method;
  therefore, we can derive $\Omega({\sqrt{N}}/{\eps_{\mathrm{min}}})$ in the quantum biased setting.
\qed
\end{proof}

\section{Estimating Unknown $\eps$}\label{se:unknown}
In Sect.\ref{se:known}, we described algorithms 
by using a given oracle $O^{\eps}_{f}$ when we know $\eps$. In this
section, we assume that there is no prior knowledge of $\eps$. 


Our overall approach is to estimate $\eps$ with appropriate accuracy
(in precise $\eps_{\mathrm{min}}$) in advance, which then can be used in  the
simulating algorithm in Lemma~\ref{le:main}. 
In the following, we first describe an overview of our strategy to
estimate $\eps_{\mathrm{min}}$ rather informally, followed by rigorous and
detailed descriptions. 

First, let us consider estimating $\eps_{x}$ in the same way as
Lemma~\ref{le:main} in quantum parallelism. 
Then, let $M^*$ denote the number of required oracle calls to achieve
a {\it good} estimation of $\eps_{x}$ for any $x$. (Here, \textit{good} means
accurate enough to perform effective amplitude amplification in Lemma~\ref{le:main}.)
Note that $M^* \in \Omega(1/\eps_{\mathrm{min}})$, and if we know the value of
$\eps$, we can set $\Theta(1/\eps)$ as $M^*$.  
However, now $\eps$ is unknown, we estimate $M^*$ as follows. 
First we will construct an algorithm, $\mathcal A_{\mathrm{enough}}$, which 
receives an input $M$ and decides whether $M$ is the number of oracle
calls to obtain a good estimation of $\eps_x$. More precisely, 
$\mathcal A_{\mathrm{enough}}$ uses $O(M)$ queries and returns $0$ if the input
$M$ is large enough to estimate $\eps_x$, otherwise it returns $1$ with
a more than constant probability, say, $9/10$. 
Then, by using $\mathcal A_{\mathrm{enough}}$ in a superposition of $\ket{x}$
as in Lemma~\ref{le:zestimation}, we can obtain the state 
$\sum_x\ket{x}\otimes\left(\alpha_x\ket{u_x}\ket{1} + \beta_x\ket{u'_x}\ket{0}\right)$.
When $M$ is small, the condition $\exists x; |\alpha_x|^2 \ge 9/10$ holds,
which means there exists $x$ such that the estimation of $\eps_x$ may be bad.
On the other hand, when $M$ is sufficiently large, the condition
$\forall x; |\alpha_x|^2 \le 1/10$ holds, which means the estimation
is good for any $x$.  Our remaining essential task, then, is to know an input
value of $M$ at the verge of the above two cases. 
Note that the value is $\Theta(1/\eps_{\mathrm{min}})$, which can be used as $M^*$.

Next, we consider an algorithm, $A_{\mathrm{check}}$,  which can distinguish the above
two cases with $O(T)$ oracle queries with a constant probability. 
Then, $M^*$ can be estimated by $O(TM^*\log\log{}M^*)$ queries by the
following search technique and majority voting:
We can find $M^*$ by trying $\mathcal A_{\mathrm{check}}$ along with exponentially increasing the
input value $M$ until $\mathcal A_{\mathrm{check}}$ succeeds.
Note that a $\log\log{}M^*$ factor is 
needed to boost the success probability of $\mathcal A_{\mathrm{check}}$ to close to 1. 
It should be noted that we cannot use robust quantum search
  algorithm~\cite{mosca} as $\mathcal A_{\mathrm{check}}$,
since there may exist $x$ such that $|\alpha_x|^2 \approx 1/2$, which
cannot be dealt with by their algorithm. 
Instead, in Lemma~\ref{le:check}, we will describe the algorithm 
$\mathcal A_{\mathrm{check}}$, which can distinguish the above two cases by 
using amplitude estimation querying for $O(\sqrt{N}\log{}N)$ times.
Then, the whole algorithm requires $O(TM^*\log\log{}M^*) =
O\left(\frac{\sqrt{N}\log{}N}{\eps_{\mathrm{min}}}\log\log{}\frac{1}{\eps_{\mathrm{min}}}\right)$ queries.
In Lemma~\ref{le:zestimation}, we present an algorithm $Par\_Est\_Zero$
that acts as $\mathcal A_{\mathrm{enough}}$ in a superposition of $\ket{x}$,
and in Lemma~\ref{le:check}, we describe the algorithm $Chk\_Amp\_Dn$ as $\mathcal A_{\mathrm{check}}$.
Finally, the whole algorithm to estimate $M^*$ is presented in Theorem~\ref{th:eminestimation}.

\begin{lemma}\label{le:zestimation}
  Let $\mathcal{O}$ be any quantum algorithm that uses no measurements such that
  $
  \mathcal{O}\ket{x}\ket{\mathbf{0}} = \ket{x}\ket{\Psi_x} = \ket{x}(\ket{\Psi_x^1} + \ket{\Psi_x^0}).
  $
  Let $\chi:\integer \to \{0,1\}$ be a Boolean function
  that divides a state $\ket{\Psi_x}$ into a good state $\ket{\Psi_x^1}$ and a bad state $\ket{\Psi_x^0}$
  such that $\sin^2(\theta_x) = \braket{\Psi_x^1}$ for any $x$ $(0 < \theta_x \leq \pi/2)$.
  There exists a quantum algorithm $Par\_Est\_Zero(\mathcal{O},\chi,M)$ that changes states as follows:
  \begin{equation*}
    \ket{x}\ket{\mathbf{0}}\ket{0} \to \ket{x}\otimes\left(\alpha_x\ket{u_x}\ket{1} + \beta_x\ket{u'_x}\ket{0}\right),
  \end{equation*}
  where $\displaystyle |\alpha_x|^2 = \frac{\sin^2(M\theta_x)}{M^2{}\sin^2(\theta_x)}$ for any $x$.
  It uses $\mathcal O$ and its inverse for $O(M)$ times.
\end{lemma}

\begin{proof}
  The algorithm $Par\_Est\_Zero(\mathcal{O},\chi,M)$ acts as $Par\_Est\_Phase(\mathcal{O},\chi,M)$
  from Step~\ref{init_state} to Step~\ref{finverse}, and applies a unitary transformation corresponding to
  the following function $g'(x)$ instead of $g_M(x)$ at Step~\ref{laststep},
  \begin{equation*}
    g'(x) = \begin{cases}
      1 & (x=0)\\
      0 & (otherwise).
    \end{cases}
  \end{equation*}
  Then, like (\ref{estimatedstate}) we have the state
  \begin{eqnarray*}
    \ket{x}&\otimes&
    \frac{-\imath}{\sqrt{2}}
    \Biggl( e^{\imath\theta_x}\ket{\Psi_x^+}
  \biggl(\alpha_{x,0}^+\ket{0}\ket{1} + \sum_{j=1}^{M-1}\alpha_{x,j}^+\ket{j}\ket{0}\biggr)\\
  &-& e^{-\imath\theta_x}\ket{\Psi_x^-}
  \biggl(\alpha_{x,0}^-\ket{0}\ket{1} + \sum_{j=1}^{M-1}\alpha_{x,j}^-\ket{j}\ket{0}\biggr)
  \Biggr),
  \end{eqnarray*}
  where
  $|\alpha_{x,j}^\pm|^2 = \frac{\sin^2(M\Delta_{x,j}^\pm\pi)}{M^2\sin^2(\Delta_{x,j}^\pm\pi)}$ such that
  $\Delta_{x,j}^+ = d(\frac{j}{M},\frac{\theta_x}{\pi})$ and $\Delta_{x,j}^- = d(\frac{j}{M},1-\frac{\theta_x}{\pi})$
  for any $x,j$. (Precisely speaking, $|\alpha_{x,j}^\pm|^2 = 1$ when $\Delta_{x,j}^\pm = 0$.
    However, $\Delta_{x,0}^\pm \neq 0$ since $\theta_x \neq 0$ in this case.)
  Note that $\ket{\Psi_x^+}$ and $\ket{\Psi_x^-}$ are mutually orthogonal and $\braket{\Psi_x^\pm} = 1$.
  Therefore, for any $x$ the last qubit has $\ket{1}$ with probability
  \[
  \frac{|\alpha_{x,0}^+|^2}{2} + \frac{|\alpha_{x,0}^-|^2}{2} = \frac{\sin^2(M\theta_x)}{M^2\sin^2(\theta_x)}.
  \]
  $Par\_Est\_Zero(\mathcal{O},\chi,M)$ requires $O(M)$ queries to $\mathcal{O}$.
  They are used when the algorithm is working as $Par\_Est\_Phase(\mathcal{O},\chi,M)$.
\qed
\end{proof}

\begin{lemma}\label{le:check}
  Let $\mathcal{O}$ be any quantum oracle such that $\mathcal{O}\ket{x}\ket{\mathbf{0}}\ket{0} =
  \ket{x} (\alpha_x\ket{w_x}\ket{1}+\beta_x\ket{u_x}\ket{0})$.
  There exists a quantum algorithm $Chk\_Amp\_Dn(\mathcal O)$ that outputs $b \in \{0,1\}$ such that
  \begin{equation*}
    b = \begin{cases}
      1 & if ~~ \exists{}x; |\alpha_x|^2 \ge \frac{9}{10}\\
      0 & if ~~ \forall{}x; |\alpha_x|^2 \le \frac{1}{10}\\
      don't ~ care & otherwise,
    \end{cases}
  \end{equation*}
  with probability at least $8/\pi^2$ using $O(\sqrt{N}\log{N})$ queries to $\mathcal O$.
\end{lemma}

\begin{proof}
  Using $O(\log{}N)$ applications of $\mathcal O$ and majority voting, we have a new oracle $\mathcal{O'}$
  such that
  \begin{eqnarray*}
    \mathcal{O'}\ket{x}\ket{\mathbf{0}}\ket{0} =
    \ket{x} (\alpha'_x\ket{w'_x}\ket{1}+\beta'_x\ket{u'_x}\ket{0}),
  \end{eqnarray*}
  where $|\alpha'_x|^2 \ge 1-\frac{1}{16N}$ if $|\alpha_x|^2 \ge \frac{9}{10}$, and
  $|\alpha'_x|^2 \le \frac{1}{16N}$ if $|\alpha_x|^2 \le \frac{1}{10}$.
  Note that work bits $\ket{w'_x}$ and $\ket{u'_x}$ are likely larger than $\ket{w_x}$ and $\ket{u_x}$.

  Now, let $\mathcal A$ be a quantum algorithm that makes
  the uniform superposition $\frac{1}{\sqrt{N}} \sum_x \ket{x}\ket{\mathbf 0}\ket{0}$ by the Fourier
  transform $\mathbf F_N$ and applies the oracle $\mathcal O'$.
  We consider (success) probability $p$ that the last qubit in the final
  state $\mathcal A\ket{\mathbf 0}$ has $\ket{1}$.
  If the given oracle $\mathcal O$ satisfies $\exists{}x; |\alpha_x|^2 \ge \frac{9}{10}$ (we call Case $1$),
  the probability $p$ is at least $\frac{1}{N} \times (1-\frac{1}{16N}) \ge \frac{15}{16N}$.
  On the other hand, if $\mathcal O$ satisfies $\forall{}x; |\alpha_x|^2 \le \frac{1}{10}$
  (we call Case $2$), then the probability
  $p \le N \times \frac{1}{N} \times \frac{1}{16N} = \frac{1}{16N}$. We can distinguish the two cases
  by amplitude estimation as follows.

  Let $\tilde\theta_p$ denote the output of the amplitude estimation
  $Est\_Phase$$(\mathcal A,\chi,\lceil 11\sqrt{N} \rceil)$.
  The whole algorithm $Chk\_Amp\_Dn$$(\mathcal O)$ performs $Est\_Phase(\mathcal A,\chi,\lceil 11\sqrt{N} \rceil)$ and
  outputs whether $\tilde \theta_p$ is greater than ${0.68}/{\sqrt{N}}$ or not.
  We will show that it is possible to distinguish the above two cases
  by the value of $\tilde\theta_p$. Let $\theta_p = \sin^{-1}(\sqrt{p})$ such that $0 \le \theta_p \le \pi/2$.
  Note that $x \le \sin^{-1}(x) \le \pi{}x/2$ if $0 \le x \le 1$.
  Theorem~\ref{th:estimation} says that in Case $1$, the $Est\_Phase$ outputs
  $\tilde\theta_p$ such that
  \[
  \tilde\theta_p \ge \theta_p - \frac{\pi}{11\sqrt{N}}
  \ge \sqrt{\frac{15}{16N}} - \frac{\pi}{11\sqrt{N}} > \frac{0.68}{\sqrt{N}},
  \]
  with probability at least $8/\pi^2$. Similarly in Case $2$, the inequality
  $\tilde\theta_p < \frac{0.68}{\sqrt{N}}$ is obtained.

  $Chk\_Amp\_Dn$$(\mathcal O)$ uses $\mathcal O$ for $O(\sqrt{N}\log{}N)$ times since
  $Chk\_Amp\_Dn$$(\mathcal O)$ calls the algorithm $\mathcal A$ for $\lceil 11\sqrt{N} \rceil$ times
  and $\mathcal A$ uses $O(\log{}N)$ queries to the given oracle $\mathcal O$.
\qed
\end{proof}

\begin{theorem}\label{th:eminestimation}
  Given a quantum biased oracle $O^{\eps}_{f}$,
  there exists a quantum algorithm $Est\_Eps\_Min(O^{\eps}_f)$ that outputs $\tilde\eps_{\mathrm{min}}$ such that
  ${\eps_{\mathrm{min}}}/{5\pi^2} \le \tilde\eps_{\mathrm{min}} \le \eps_{\mathrm{min}}$ with probability at least $2/3$.
  The query complexity of the algorithm is expected to be
  $
  O\left(\frac{\sqrt{N}\log{N}}{\eps_{\mathrm{min}}} \log\log\frac{1}{\eps_{\mathrm{min}}}\right).
  $
\end{theorem}


\begin{proof}
  Let $\sin(\theta_x) = 2\eps_x$ and $\sin(\theta_{\mathrm{min}}) = 2\eps_{\mathrm{min}}$ such that
  $0<\theta_x,\theta_{\mathrm{min}}\le\frac{\pi}{2}$. Let
  $\chi$ also be a Boolean function that divides the state in (\ref{eq:oracle})
  into a good state $(-1)^{f(x)}2\eps_x\ket{0^{m+1}}$ and a bad state $\ket{\psi_x}$.
  Thus $Par\_Est\_Zero(\tilde{O}_{f}^{\eps},\chi,M)$ in Lemma~\ref{le:zestimation} makes the state
  $\ket{x}\otimes\left(\alpha_x\ket{u_x}\ket{1} + \beta_x\ket{u'_x}\ket{0}\right)$
  such that $|\alpha_x|^2 = \frac{\sin^2(M\theta_x)}{M^2\sin^2(\theta_x)}$.
  As stated below, if $M \in o(1/\theta_x)$, then $|\alpha_x|^2 \ge 9/10$.
  We can use $Chk\_Amp\_Dn$ to check whether there exists $x$ such that $|\alpha_x|^2 \ge 9/10$.
  Based on these facts, we present the whole algorithm $Est\_Eps\_Min(O^{\eps}_f)$.
  
\begin{algorithm}{$Est\_Eps\_Min(O^{\eps}_f)$}
  \begin{enumerate}
  \item Start with $\ell=0$.
  \item\label{inc} Increase $\ell$ by $1$.
  \item\label{branch} Run $Chk\_Amp\_Dn(Par\_Est\_Zero(\tilde{O}_{f}^{\eps},\chi,2^\ell))$
    for $O(\log\ell)$ times and use majority voting.
    If ``$1$'' is output as the result of the majority voting, then return to Step~\ref{inc}.
  \item Output $\tilde\eps_{\mathrm{min}} = \frac{1}{2}\sin\left(\frac{1}{5\cdot{}2^\ell}\right)$.
  \end{enumerate}
\end{algorithm}
  
  Now, we will show that the algorithm almost keeps running until
  $\ell > \left\lfloor{\log_2\frac{1}{5\theta_{\mathrm{min}}}}\right\rfloor$.
  We assume $\ell \leq \left\lfloor{\log_2\frac{1}{5\theta_{\mathrm{min}}}}\right\rfloor$. Under this assumption,
  a proposition $\exists x; |\alpha_x|^2 \geq \frac{9}{10}$ holds since the equation
  $\eps_{\mathrm{min}} = \min_x{\eps_x}$ guarantees that there exists some $x$ such that $\theta_{\mathrm{min}} = \theta_x$
  and $|\alpha_x|^2 = \frac{\sin^2(2^\ell\theta_x)}{2^{2\ell}\sin^2(\theta_x)}
  \geq \cos^2(\frac{1}{5}) > \frac{9}{10}$ when $2^\ell \leq \frac{1}{5\theta_x}$.
  Therefore, a single $Chk\_Amp\_Dn$ run returns ``$1$'' with probability at least $8/\pi^2$.
  By $O(\log\ell)$ repetitions and majority voting, the probability that we obtain ``$1$'' increases to
  at least $1 - \frac{1}{5\ell^2}$.
  Consequently, the overall probability
  that we return from Step~\ref{branch} to Step~\ref{inc} for any $\ell$ such that
  $\ell \leq \left\lfloor{\log_2\frac{1}{5\theta_{\mathrm{min}}}}\right\rfloor$ is at least
  $
  \prod_{\ell=1}^{\left\lfloor{\log_2\frac{1}{5\theta_{\mathrm{min}}}}\right\rfloor} \left(1-\frac{1}{5\ell^2}\right)
  > \frac{2}{3}.
  $ This inequality can be obtained by considering an infinite product expansion of $\sin(x)$,
  i.e., $\sin(x) = x\prod_{n=1}^{\infty}\left(1-\frac{x^2}{n^2\pi^2}\right)$ at $x=\pi/\sqrt{5}$.
  Thus the algorithm keeps running until $\ell > \left\lfloor{\log_2\frac{1}{5\theta_{\mathrm{min}}}}\right\rfloor$,
  i.e., outputs $\tilde\eps_{\mathrm{min}}$ such that
  $\tilde\eps_{\mathrm{min}} = \frac{1}{2}\sin\left(\frac{1}{5\cdot{}2^\ell}\right) \le \frac{1}{2}\sin(\theta_{\mathrm{min}}) = \eps_{\mathrm{min}}$,
  with probability at least $2/3$.
  
  We can also show that the algorithm almost stops in $\ell < \left\lceil\log_2\frac{2\pi}{\theta_{\mathrm{min}}}\right\rceil$.
  Since $\frac{\sin^2(M\theta)}{M^2\sin^2(\theta)}
  \leq \frac{\pi^2}{(2M\theta)^2}$ when $0 \le \theta \le \frac{\pi}{2}$,
  $|\alpha_x|^2 = \frac{\sin^2(2^\ell\theta_x)}{2^{2\ell}\sin^2(\theta_x)} \leq \frac{1}{16}$ for any $x$
  if $2^\ell \geq \frac{2\pi}{\theta_{\mathrm{min}}}$.
  Therefore, in Step~\ref{branch}, ``$0$'' is returned with probability at least ${8}/{\pi^2}$
  when $\ell \geq \left\lceil\log_2\frac{2\pi}{\theta_{\mathrm{min}}}\right\rceil$.
  The algorithm, thus, outputs
  $\tilde\eps_{\mathrm{min}} = \frac{1}{2}\sin\left(\frac{1}{5\cdot{}2^\ell}\right)
  \ge \frac{1}{2}\sin(\frac{\theta_{\mathrm{min}}}{10\pi}) \ge \frac{\eps_{\mathrm{min}}}{5\pi^2}$
  with probability at least ${8}/{\pi^2}$.
  
  Let $\tilde\ell$ satisfy 
  $\left\lfloor{\log_2\frac{1}{5\theta_{\mathrm{min}}}}\right\rfloor <\tilde\ell
  < \left\lceil\log_2\frac{2\pi}{\theta_{\mathrm{min}}}\right\rceil$.
  If the algorithm runs until $\ell= \tilde\ell$,
  its query complexity is
  \begin{eqnarray*}
    \sum_{\ell = 1}^{\tilde{\ell}} O(2^\ell{}\sqrt{N}\log N\log\ell)
    = O(2^{\tilde\ell}\sqrt{N}\log{N}\log{\tilde\ell})
    = O\left(\frac{\sqrt{N}\log{N}}{\eps_{\mathrm{min}}}\log\log\frac{1}{\eps_{\mathrm{min}}}\right),
  \end{eqnarray*}
  since $2^{\tilde\ell}
  \in \Theta\left(\frac{1}{\theta_{\mathrm{min}}}\right) = \Theta\left(\frac{1}{\eps_{\mathrm{min}}}\right)$.
\qed
\end{proof}

\noindent                                                                               
{\bf Remark.~} 
As mentioned above, we have some way to deal with quantum biased oracles even if
we have no knowledge about the given oracle's bias rate.
On the other hand, in the classical biased setting, there seems to be no
way if the value of $\eps$ is unknown: Suppose that classical biased
oracles return a correct value with probability at least $1/2+\eps$ for each query.
It is known that by using $O(1/\eps^2)$ queries and majority voting,
the probability that oracles answer queries correctly increases to $2/3$. 
However, this algorithm works effectively when we know $\eps$. In
other words, unless we know $\eps$, it is likely impossible to
determine an appropriate number of majority voting to achieve at least 
a constant success probability of the whole algorithm.  

\section{Conclusion}
In this paper, we have shown that $O(\sqrt{N}/{\eps})$ queries are
enough to compute $N$-bit OR with an ${\eps}$-biased oracle. 
This matches the known lower bound while affirmatively answering the conjecture raised 
by the paper~\cite{yamashita}. 
The result in this paper implies other matching bounds such as computing parity
with $\Theta(N/\eps)$ queries.
We also show a quantum algorithm that estimates unknown value of
${\eps}$ with an ${\eps}$-biased oracle. Then, by using the estimated
value, we can construct a robust algorithm even when $\eps$ is
unknown. This contrasts with the corresponding classical case where no
good estimation method seems to exist.    

 Until now, unfortunately, we have had essentially only one quantum
 algorithm, i.e., the robust quantum search algorithm~\cite{mosca}, 
 to cope with imperfect oracles. (Note that other algorithms, including 
 our own algorithm in Theorem~\ref{th:or}, are all based on the robust
 quantum search algorithm~\cite{mosca}.)   
 Thus, it should be interesting to seek another {\it essentially  
  different} quantum algorithm with imperfect oracles.  
 If we find a new quantum algorithm that uses $O(T)$ queries to 
 imperfect oracles with constant probability, then we can have
 a quantum algorithm that uses $O(T/\eps)$ queries to imperfect oracles
 with an ${\eps}$-biased oracle based on our method. This is different
 from the classical case where we need an overhead factor of $O(1/{\eps}^2)$
 by majority voting.

\section*{Appendix}
Here, we describe the algorithm and the proof of Lemma~\ref{le:pestimation} after providing a few definitions.
The algorithm $Par\_Est\_Phase(\mathcal{O},\chi,M)$ is based on the
amplitude estimation algorithm in \cite{qaa}. 
We refer interested readers to \cite{qaa}.

\begin{definition}\label{distance}
For any two real numbers $\omega_0, \omega_1 \in \reals$,
\[
\displaystyle d(\omega_0,\omega_1)
= \min_{z \in \integer}\{|z+\omega_1-\omega_0| \}.
\]
\end{definition}

Thus $2\pi d(\omega_0,\omega_1)$ is the length of the shortest arc
on the unit circle going from $e^{2\pi \imath \omega_0}$
to $e^{2 \pi \imath \omega_1}$.
Note that $0 \le d(\omega_0,\omega_1) \le \frac{1}{2}$ for any $\omega_0,\omega_1$.

\begin{definition}
  For any integer $M \geq 1$, let $g_M(x)$ be a function defined by
  \begin{equation*}
    g_M(x) = \begin{cases}
      \frac{\pi{}x}{M} & (0\le x \le \frac{M}{2})\\
      \pi - \frac{\pi{}x}{M} & (\frac{M}{2} \le x < M).
    \end{cases}
  \end{equation*}
\end{definition}

\begin{algorithm}{$Par\_Est\_Phase(\mathcal{O},\chi,M)$}
  \begin{enumerate}
  \item\label{init_state} Start with
    the state $\ket{x}\ket{\mathbf{0}}\ket{\mathbf 0}\ket{\mathbf 0}$.
  \item\label{1query} Apply $\mathcal O$ to the first and the second registers.
  \item\label{est_init}Apply ${\mathbf F}_M$ to the third register.
  \item\label{est_lambda} Apply $\Lambda_M(\mathbf Q)$ controlled by the third register,
    where $\mathbf Q = -\mathcal{O}(\mathbf I \otimes \mathbf S_{0})
    {\mathcal{O}}^{-1} (\mathbf I \otimes \mathbf S_{\chi})$.
    $\mathbf Q$ is applied to the first and the second registers.
  \item\label{finverse} Apply ${\mathbf F}_M^{-1}$ to the third register.
  \item\label{laststep} Apply the unitary transformation
    $\mathbf U_{g_M}$ to the third and the fourth registers,
    where $\mathbf U_{g_M}$ maps $\ket{x}\ket{\mathbf{0}}$ to $\ket{x}\ket{g_M(x)}$.
  \end{enumerate}
\end{algorithm}

\begin{proof}(of Lemma~\ref{le:pestimation})
  
  When $\theta_x = 0,\frac{\pi}{2}$, the analysis can be performed almost like the following;
  therefore, we assume $0<\theta_x< \frac{\pi}{2}$ for any $x$.
  Focusing on the subspace where the first register has a basis state $\ket{x}$,
  the transformation $\mathbf Q_x^j{}\mathcal O_x$ is applied to the second register, where
  $\mathbf Q_x = -\mathcal O_x\mathbf S_0\mathcal O_x^{-1}\mathbf S_\chi$
  and $j$ is the number designated by the
  third register. Since this situation is the same as
  in Theorem 12 in \cite{qaa}, the analysis can be done similarly.
  Let
  \[
  \ket{\Psi_x^\pm} =
  \frac{1}{\sqrt{2}}\left(\frac{1}{\sin\theta_x}\ket{\Psi_x^1}\pm\frac{\imath}{\cos\theta_x}\ket{\Psi_x^0}\right).
  \]
  Note that $\ket{\Psi_x^+}$ and $\ket{\Psi_x^-}$ are orthonormal eigenvectors of $\mathbf Q_x$.
  After Step~\ref{laststep}, we can obtain the state
  \begin{eqnarray}
    \ket{x}&\otimes&
    \frac{-\imath}{\sqrt{2}}
    \Biggl( e^{\imath\theta_x}\ket{\Psi_x^+}
    \biggl(\sum_{j=0}^{M-1}\alpha_{x,j}^+\ket{j}\ket{g_M(j)}\biggr) \nonumber \\
    &-& e^{-\imath\theta_x}\ket{\Psi_x^-}
    \biggl(\sum_{j=0}^{M-1}\alpha_{x,j}^-\ket{j}\ket{g_M(j)}\biggr) \label{estimatedstate}
    \Biggr),
  \end{eqnarray}
  where
  $|\alpha_{x,j}^\pm|^2 = \frac{\sin^2(M\Delta_{x,j}^\pm\pi)}
  {M^2\sin^2(\Delta_{x,j}^\pm\pi)}$ such that
  $\Delta_{x,j}^+ = d(\frac{j}{M},\frac{\theta_x}{\pi})$
  and $\Delta_{x,j}^- = d(\frac{j}{M},1-\frac{\theta_x}{\pi})$
  for any $x,j$. (Precisely speaking, $|\alpha_{x,j}^\pm|^2 = 1$ when $\Delta_{x,j}^\pm = 0$.
    This condition means that $\frac{M\theta_x}{\pi}$ or $M-\frac{M\theta_x}{\pi}$ is an integer.)
  This follows Theorem 11 in \cite{qaa}.

  We will show that the last register has the good estimations of $\theta_x$ with high probability in the final state.
  Now, let $j_1^+ = \lfloor\frac{M\theta_x}{\pi}\rfloor$ and $j_2^+ = \lceil\frac{M\theta_x}{\pi}\rceil$.
  $0 < \theta_x < \frac{\pi}{2}$ means $0 \le j_i^+ \le \frac{M}{2}$, thus
  $g_M(j_i^+) = \frac{j_i^+\pi}{M}$ holds. We can prove
  $|\alpha_{x,j_1^+}^+|^2 + |\alpha_{x,j_2^+}^+|^2 \geq \frac{8}{\pi^2}$ and
  $|\theta_x - g_M(j_i^+)| \leq \frac{\pi}{M}$ like Theorem 11 in \cite{qaa}.
  Similarly, let $j_1^- = M - \lfloor\frac{M\theta_x}{\pi}\rfloor$ and $j_2^- = M - \lceil\frac{M\theta_x}{\pi}\rceil$.
  $0 < \theta_x < \frac{\pi}{2}$ means $\frac{M}{2} \le j_i^- \le M$, thus
  $g_M(j_i^-) = \pi - \frac{j_i^-\pi}{M}$ holds. (This holds when $\frac{M}{2} \le j_i^- < M$.
  When $j_i^- = M$, we consider $j_i^-=0$, then the following sentences will hold.)
  We can also prove
  $|\alpha_{x,j_1^-}^-|^2 + |\alpha_{x,j_2^-}^-|^2 \geq \frac{8}{\pi^2}$ and
  $|\theta_x - g_M(j_i^-)| \leq \frac{\pi}{M}$.
  Thus the probability that the last register \ket{g_M(j)} has an estimated value $\tilde\theta_x$
  such that $|\theta_x - \tilde\theta_x| \leq \frac{\pi}{M}$ is
  \begin{eqnarray*}
    && \sum_{j:|\theta_x - g_M(j)|\leq \frac{\pi}{M}}\frac{|\alpha_{x,j}^+|^2}{2}
    + \sum_{j:|\theta_x - g_M(j)|\leq \frac{\pi}{M}}\frac{|\alpha_{x,j}^-|^2}{2}\\
    & \geq &
    \sum_{j \in \{j_1^+,j_2^+\}}\frac{|\alpha_{x,j}^+|^2}{2}
    + \sum_{j \in \{j_1^-,j_2^-\}}\frac{|\alpha_{x,j}^-|^2}{2}
    ~ \geq ~
    \frac{8}{\pi^2}.
  \end{eqnarray*}  
  Therefore, the well-estimated values of $\theta_x$ lie
  in the last register with probability at least ${8}/{\pi^2}$.
\end{proof}

\end{document}